\documentclass[letterpaper, twocolumn, prl, showpacs]{revtex4}
\usepackage[T1]{fontenc}
\usepackage{graphics, graphicx, amsmath, pifont}

\begin{document}
\newcommand{\zl}[2]{$#1\:\text{#2}$}
\newcommand{\zlm}[2]{$#1\:\mu\text{#2}$}
\title{Single-Photon Atomic Cooling}
\date{December 28, 2007}
\author{Gabriel N. Price}
\author{S. Travis Bannerman}
\author{Kirsten Viering}
\author{Edvardas Narevicius}
\author{Mark G. Raizen}
\email{raizen@physics.utexas.edu}
\affiliation{Center for Nonlinear Dynamics and Department of Physics, The University of Texas at Austin, Austin, Texas 78712, USA}

\begin{abstract}
We report the cooling of an atomic ensemble with light, where each atom scatters only a \textit{single photon on average}. This is a general method that does not require a cycling transition and can be applied to atoms or molecules which are magnetically trapped. We discuss the application of this new approach to the cooling of hydrogenic atoms for the purpose of precision spectroscopy and fundamental tests.
\end{abstract}

\maketitle
Cooling and trapping of atoms in the gas phase has been a central theme in physics for over thirty years \cite{metcalf}.  The main advances in this field were enabled by laser cooling, which relies on the transfer of momentum from photons to atoms in a cycle of repeated scattering.  Despite the enormous success of this method, it has been limited to a small set of atoms in the periodic table due to the need for a two-level cycling transition that is accessible with stabilized lasers.\par
We have been working to develop more general methods to trap and cool atoms which would be applicable to most of the periodic table as well as to many molecules.  Our approach has been to divide the task into two parts.  The first is to stop a supersonic atomic or molecular beam with pulsed magnetic fields as reported in Ref. \cite{stopping supersonic beams}.  That step provides an atomic or molecular sample which is magnetically trapped at temperatures in the \zl{10}{mK} range.  The second step, reported in this Letter, is to develop a method that can further cool the atoms or molecules but does not require a cycling transition.  An existing, non-laser based method is evaporative cooling, which has been successfully employed to reach Bose-Einstein condensation \cite{ketterle}.  However, this approach is even more restrictive than laser cooling due to the severe constraints on the nature of the interparticle collisions.  We report here on a new approach that can accumulate atoms or molecules from a magnetic trap into an optical dipole trap.  The method is based on the concept of a "one-way wall of light" for atoms and molecules that was introduced in a series of earlier publications \cite{raizen, ar, dudarev, kim, muga}.  The experimental realization of this principle is presented here and sets a general framework for cooling.\par
We implement single-photon atomic cooling for the specific case of $^{87}$Rb, using a scheme similar to one previously proposed by our group \cite{price}. This scheme transfers atoms from a large-volume magnetic trap into a small-volume optical trap via an irreversible optical pumping step which requires each atom to scatter only one photon. By loading from the wing of the magnetic trap, we selectively transfer only atoms near their classical turning points where they have little kinetic energy. As the outer shell of the magnetic trap is depleted, we adiabatically translate the trap center toward the optical trap for maximum loading and phase-space compression.\par
The experimental apparatus is similar to that described in previous work \cite{meyrath}. A thermal cloud of $^{87}$Rb atoms is initially produced in a magneto-optical trap and then cooled in optical molasses. Subsequently atoms in the $5 S_{1/2}(F=2)$ hyperfine ground state are loaded into a magnetic quadrupole trap with a radial field gradient of \zl{75}{G/cm}. We trap approximately $1.7 \times 10^8$ atoms at a temperature of \zlm{90}{K} in a cloud with a $1/e$ radius of \zlm{550}{m}.\par
\begin{figure}
\includegraphics[width=.4\textwidth]{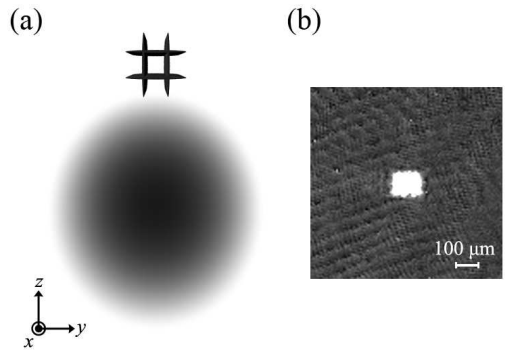}
\caption{\label{fig1} (a) Cross-section of the optical box positioned above the cloud of magnetically trapped atoms. In this illustration, two pairs of Gaussian laser sheets propagate parallel to the x-axis. A third pair (not visible) propagates parallel to the y-axis and completes the optical box. (b) Absorption image along the z-axis of approximately $1.5 \times 10^5$ atoms trapped in the optical box.}
\end{figure}

 
After the magnetic trap is loaded, an optical dipole trap is positioned above it. The optical dipole trap originates from a single-mode \zl{10}{W} laser at \zl{\lambda=532}{nm} which is split into three beams. Each beam passes through a dual-frequency acousto-optic modulator, and the first order deflections are tightly focused in one dimension to form parallel sheets. Each individual sheet has a $1/e^2$ beam waist of \zlm{10}{m} \zlm{\times \; 200}{m} and a power of \zl{0.7}{W}.  The three pairs of sheets are crossed to form a repulsive "box-like" potential, with dimensions \zlm{100}{m} \zlm{\times \; 100}{m} \zlm{\times \; 130}{m} and a depth of \zlm{k_B \times 10}{K}, shown pictorially in Fig.\ref{fig1} (a).\par  
The accumulation of atoms in the optical box, a conservative trap, requires an irreversible step. This need is met by optically pumping the atoms that transit the optical box to the $F=1$ manifold with a so-called depopulation beam. The beam is resonant with the $5S_{1/2} (F=2) \rightarrow 5P_{3/2} (F'=1)$ transition and focused to a $1/e^2$ waist of \zlm{8}{m} \zlm{\times \; 200}{m} at the center of the box.  Magnetically trapped atoms in the $F=2$ manifold are excited by the depopulation beam and decay with $84\%$ probability to the $F=1$ manifold $(m_F=1,0)$, where they are no longer on resonance with the depopulation beam. Because the gradient of the Zeeman shift of these states is smaller than that of the initial state, the contribution from the magnetic field to the total potential is reduced, creating a trapped state in the optical box \cite{price}.\par 
As atoms accumulate in the optical box, the outermost trajectories of the magnetic trap are depleted by the depopulation beam. For maximum loading into the optical box, we adiabatically translate the center of the magnetic trap towards the optical box by applying a linear current ramp to an auxiliary magnetic coil located above the atoms.\par
Before imaging, we isolate the optically trapped atoms by switching off the magnetic trap, allowing untrapped atoms to fall under the influence of gravity for \zl{80}{ms}. Additionally, the depopulation beam is turned off and a beam resonant with the $5S_{1/2}(F=2) \rightarrow 5P_{3/2}(F'=3)$ transition blows away any residual atoms in the $F=2$ manifold. The remaining atoms are those which have undergone single-photon atomic cooling. These atoms are pumped to the $F=2$ manifold and illuminated with freezing molasses for \zl{30}{ms}. The resulting fluorescence is imaged on a charge-coupled device (CCD) camera and integrated to yield atom number. Spatial information is obtained by imaging with absorption rather than fluorescence as in Fig. \ref{fig1} (b).\par
The density of atoms loaded into the optical box via single-photon atomic cooling is sensitive to multiple parameters.  The intensity of the depopulation beam strongly affects the final density; it must be set to balance efficient pumping into the $F=1$ manifold with trap loss due to heating.  In our experimental configuration, we maximize density in the optical box with a peak depopulation beam intensity of approximately \zl{8}{mW/cm$^2$}. \par
In addition to the depopulation beam intensity, transfer into the optical box is highly affected by both the duration and range over which the magnetic trap is translated.  The optimal duration of this translation is mainly dependent on two competing factors.  Long translation times permit phase-space exploration by atoms in the magnetic trap, allowing a more complete exchange of kinetic for potential energy before an atom encounters the optical box.  However, the finite lifetime of atoms in the optical box (\zl{\tau=3.7\pm 0.1}{s} in the presence of the depopulation beam) limits the translation time. We achieve highest density with a translation time of approximately \zl{1.2}{s}.   Given this time scale, the translation range loading the largest atom number into the optical box is empirically determined. We translate the optical box from an initial separation (relative to the center of the magnetic trap) of \zlm{700}{m} to a final separation of \zlm{100}{m}.\par
\begin{figure}
\includegraphics[width=.47\textwidth]{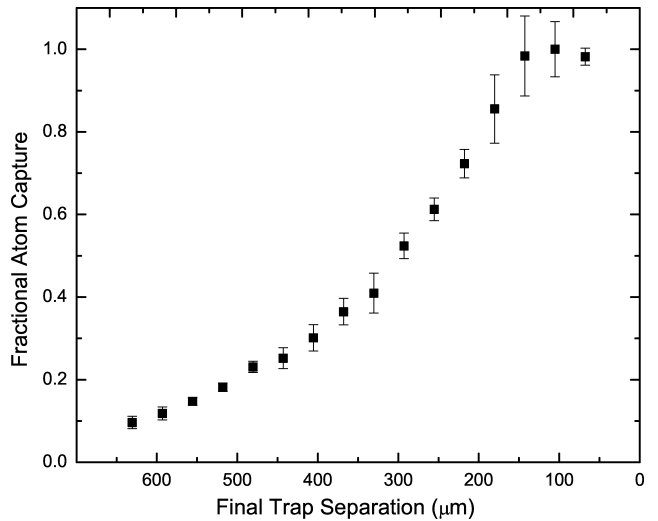}
\caption{\label{fig2} Incremental atom capture at a fixed translation velocity. The center of the magnetic trap is initially displaced \zlm{800}{m} below the optical box and is translated vertically at a velocity of \zlm{750}{m/s}. The endpoint of the translation is varied, and the atom capture, normalized to the maximum number, is plotted as a function of the final separation between the traps. Error bars indicate statistical uncertainties.} 
\end{figure}

To study the dynamics of the loading process, we look at the incremental loading for a constant translation velocity.  We start with the center of the magnetic trap \zlm{800}{m} below the optical box and then translate it vertically at a velocity of \zlm{750}{m/s}.  Figure \ref{fig2} displays the fraction of atoms captured as a function of the final separation between the magnetic trap and the optical box.  The slope of this plot indicates that the local loading rate increases with decreasing separation until about \zlm{100}{m}.  Additionally, it is clear from this plot that atom capture is not increased by translating beyond this point.\par
\begin{figure}
\includegraphics[width=.47\textwidth]{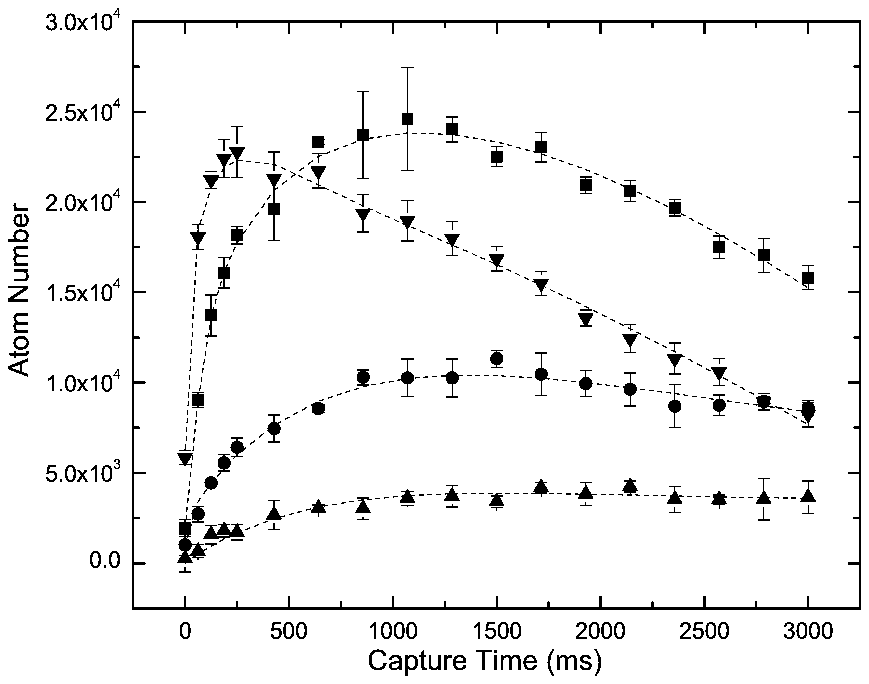}
\caption{\label{fig3}Captured atom number as a function of loading time. Data are given for separations between the optical and magnetic trap centers of \zlm{800}{m} (\ding{115}), \zlm{600}{m} (\ding{108}), \zlm{400}{m} (\ding{110}), and \zlm{200}{m} (\ding{116}). Error bars indicate statistical uncertainty, and dashed curves are drawn through the data points to guide the eye. The slopes are initially dominated by the loading rate into the optical box. After some time, the loading rate decreases due to the depletion of the magnetic trap, and the slopes become dominated by escape out of the box.} 
\end{figure}

We study loading without translating the box (i.e. at a fixed separation) to understand the dynamics of single-photon atomic cooling in more detail.  Figure \ref{fig3} shows the number of atoms loaded into the optical box as a function of time for several separations.  All curves exhibit a positive initial slope indicative of the loading rate. As the magnetic trap is depleted by the depopulation beam, the loading rate decreases and the slope becomes dominated by trap losses. We find both the loading rate and the trap loss rate to be inversely related to the separation between the magnetic trap and optical box centers.  The former reflects the dependence of the loading rate on the local density of magnetically trapped atoms.  The latter suggests a higher rate of escape out of the optical box for smaller separations.
This may be attributed in part to an increased temperature caused by collisions between atoms in the optical box and atoms in the magnetic trap.  For the two smallest separations (\zlm{200}{m}, \zlm{400}{m}) we calculate initial collision rates of (\zl{0.8}{Hz}, \zl{0.5}{Hz}) respectively. However, these rates diminish as the depopulation beam reduces the density of magnetically trapped atoms in the vicinity of the optical box. We thus consider collisions non-negligible for t < (\zl{250}{ms}, \zl{500}{ms}) which provides an upper bound of ($0.2$, $0.25$) collisions per atom in the optical box. A large fraction of these collisions will cause immediate trap loss on account of the shallow box depth (\zlm{~10}{K}), but a few will raise the temperature. We believe, however, that this effect is overshadowed by atoms entering the optical box far from their classical turning points. In contrast to adiabatically translating the magnetically trapped atoms toward the optical box (as in Fig. \ref{fig2}), which yields a kinetic energy distribution independent of translation endpoint, we abruptly turn on the optical box and depopulation beam for the data in Fig. \ref{fig3}. In this situation, many atoms now transit the optical box far from their classical turning points, and if captured they contribute to an increased kinetic energy distribution and rate of escape.\par
We performed Monte-Carlo simulations of the dynamics in the magnetic trap and transfer into the optical box. Atom trajectories are propagated through phase-space, in which a subspace representing trapped states in the optical box has been defined. As atoms reach this subspace they are counted as trapped. These simulations show an inverse relationship between the loading rate and the separation between the magnetic trap and optical box centers in agreement with the experimental results. We will present more detailed studies and quantitative comparisons in a future publication.\par 
Of utmost importance to the utility of this cooling technique is its ability to compress phase-space.  With the single-photon atomic cooling scheme described in this Letter, we extract $1.5 \times 10^5$ atoms at a temperature of \zlm{7}{K} from the magnetic trap. We compare this with the number of atoms captured out of the magnetic trap without the depopulation beam. This is just a conservative dipole trap: atoms that are caught inside the box at low enough kinetic energy will be trapped, while all others will be lost. We measure a factor of $23 \pm 3$ increase in atom number using the single-photon atomic cooling method with nearly identical velocity distributions.  We do not resolve the internal magnetic states in our measurement.  The atoms in the magnetic trap are in the $F=2$ manifold, but can be in the $m_F=1$ and $m_F=2$ magnetic sublevels.  The atoms caught in the optical box are in the $F=1$ manifold but can be in the $m_F=1$ and $m_F=0$ magnetic sublevels.  The factor of $23$ refers to atom number, not directly to phase space density.  The increase in the latter would be a factor of $12$ in the worst case scenario, if all the atoms in the magnetic trap were in the $F=2$, $m_F=2$ state and the atoms in the dipole trap were equally distributed between the two magnetic sublevels. \par
The increase in phase-space density demonstrated here is limited by technical constraints and does not represent a fundamental limit to this process. Future work is aimed at increasing the lifetime of the magnetic trap in the presence of the depopulation beam. One possible technique employs a \zl{778}{nm} depopulation beam resonant with the $^{87}$Rb $5S_{1/2} \rightarrow 5D_{5/2}$ two-photon transition \cite{snadden}. Such transitions depend more strongly on beam intensity than single-photon transitions, allowing better localization of the depopulation transition to within the confines of the optical box.\par
We emphasize that the method of single-photon atomic and molecular cooling does not rely on photon momentum transfer.  Instead, the scattering of a photon causes an irreversible change in the effective potential that traps the particle.  We showed in an earlier publication that the scattering of a photon by each atom entering the trap raises the entropy of the radiation field by an amount exactly equal to the reduction of entropy of the atoms \cite{ruschhaupt}.  In that regard, our method is informational cooling in the same sense first proposed by L. Szilard in 1929 in order to resolve the paradox of Maxwell's demon \cite{szilard, leff}.  However, unlike the demon, our method does not require actual measurement and feedback \cite{scully}, and it is maximally efficient in the sense that only one photon per atom is required.  The quantum limits of our method are still not clear, and further work is required, both in theory and experiment.  Cooling to quantum degeneracy may be possible with the atoms near the single-photon recoil temperature.\par
It is interesting to compare single-photon atomic cooling with forced RF evaporative cooling. The latter method truncates the velocity distribution with an RF knife while the former truncates the velocity distribution with the depopulation beam \cite{dudarev}. However, in contrast to forced RF evaporative cooling the ejected atoms are captured instead of lost. On the time scale of our experiment, the collision rate in the magnetic trap is not sufficient to thermalize the system, and the experiment thus proceeds with the system out of thermal equilibrium where the velocity distribution of the cloud is not independent of position.  This is in contrast to evaporative cooling, where the system must be allowed to return to near-thermal equilibrium via elastic interparticle scattering.  As a final comparison, we note that while rethermalization via many two-body elastic collisions is a collective process, single-photon atomic cooling is fundamentally a single-particle process.\par
Beyond a first demonstration experiment, the real significance of our method is that it can be applied quite generally to atoms and molecules which can be magnetically trapped.  We will apply it to the trapping and cooling of atomic hydrogen, which has been the "Rosetta Stone" of physics for many years and is the simplest and most abundant atom in the universe.  Precision spectroscopy of the hydrogen isotopes, deuterium and tritium, continues to be of great interest to atomic and nuclear physics.  Tritium is the simplest radioactive element and serves as an ideal system for the study of beta decay.  The latter may be the only way to determine the neutrino rest mass, one of the most pressing questions in contemporary physics.  Despite these very important features, hydrogen has remained very difficult to control and trap, while deuterium and tritium have never been trapped. This will be accomplished with an atomic coilgun where hydrogenic atoms will be entrained in a supersonic beam of helium \cite{stopping supersonic beams}. \par
After magnetic trapping, further cooling can be accomplished by the implementation of a single-photon atomic cooling scheme very similar to that reported for rubidium in this paper.  The $1S$ ground state of hydrogenic atoms is split into two hyperfine states, $F=0$ and $F=1$, separated by \zl{1.42}{GHz}.  Atoms can be magnetically trapped in the low-field seeking state, $F=1$, $m_F=1$.  The atoms can then be transferred to an optical dipole trap with a depopulation beam tuned to the two-photon transition at \zl{243}{nm}. This drives a transition to the $2S$ state which can then be quenched with a microwave field, followed by the spontaneous emission of a Lyman alpha photon at \zl{121}{nm}\cite{cesar}.  Atoms that decay into the $F=0$, $m_F=0$ state would be trapped.  The ideal configuration would employ an optical dipole trap tuned to a magic wavelength for the $1S$ to $2S$ transition, as that would enable spectroscopy of unprecedented precision.  In fact, a magic wavelength for this case has been predicted near \zl{515}{nm} \cite{kielpinski}, and a resonant build-up cavity could provide a trap that is a few hundred microkelvin deep.  The same method could also be used to accumulate anti-hydrogen atoms in an optical trap, enabling precise spectroscopy and a search for CPT violation \cite{alpha, atrap}.\par
Another important application of our method is the cooling of molecules \cite{price}, which will be discussed in more detail in a forthcoming paper.\par
We would like to thank M. O. Scully for insightful discussions.
We acknowledge support from the R. A. Welch Foundation, the Sid W. Richardson Foundation, and the National Science Foundation.\par

\end{document}